# Magnetic-Field-Driven Insulator-Superconductor Transition in Rhombohedral Graphene

Jian Xie[1†], Zihao Huo[1†]*, Zhimou Chen[1], Zaizhe Zhang[1], Kenji Watanabe[2], Takashi Taniguchi[3], Xi Lin[1,4,5]* and Xiaobo Lu[1,6]*

[1]*International Center for Quantum Materials, School of Physics, Peking University, Beijing 100871, China*
[2]*Research Center for Electronic and Optical Materials, National Institute for Material Sciences, 1-1 Namiki, Tsukuba 305-0044, Japan*
[3]*Research Center for Materials Nanoarchitectonics, National Institute for Material Sciences, 1-1 Namiki, Tsukuba 305-0044, Japan*
[4]*Hefei National Laboratory, Hefei 230088, China*
[5]*Interdisciplinary Institute of Light-Element Quantum Materials and Research Center for Light-Element Advanced Materials, Peking University, Beijing 100871, China*
[6]*Collaborative Innovation Center of Quantum Matter, Beijing 100871, China*

[†]These authors contributed equally to this work.
*E-mails: xiaobolu@pku.edu.cn; xilin@pku.edu.cn; zhhuo@stu.pku.edu.cn

**Recent studies of rhombohedral multilayer graphene (RMG) have revealed a variety of superconducting states that can be induced or enhanced by magnetic fields, reinforcing RMG as a powerful platform for investigating novel superconductivity. Here we report an insulator-superconductor transition driven by in-plane magnetic fields $B_\parallel$ in rhombohedral hexalayer graphene. The upper critical field of $B_\parallel \approx 2$ T violates the Pauli limit, and an analysis based on isospin symmetry breaking supports a spin-polarized superconductor. At $B_\parallel = 0$, such spin-polarized superconductor transitions into an insulator, exhibiting a thermally activated gap of $\Delta \approx 0.1$ meV. In addition, we observe four superconducting states in the hole-doped regime, as well as phases with orbital multiferroicity near charge neutrality point. These findings substantially enrich the phase diagram of rhombohedral graphene and provide new insight into the microscopic mechanisms of superconductivity.**

Within the conventional Bardeen-Cooper-Schrieffer (BCS) framework [1,2], most known superconductivity arises from spin-singlet pairing and is therefore expected to be suppressed by magnetic fields [3]. Superconductor-insulator/metal transitions (SIT/SMT) in two-dimensional superconductivity typically result from the suppression of the amplitude or the loss of long-range phase coherence of superconducting order parameter [4–7]. In this picture, magnetic fields and superconductivity are fundamentally antagonistic. In contrast, spin-polarized (spin-triplet) superconducting states exhibit remarkable robustness or even preference under magnetic fields, featuring Pauli limit violation. Theoretical and experimental investigations of spin-polarized superconducting remain crucial.

Graphene-based flat-band systems provide a particularly favorable setting for exploring spin-polarized superconductivity. Experimentally, besides systems such as Uranium compounds [8–15] and perovskite [16–18], signatures of spin-polarized superconductivity have been frequently observed in graphene-based structures, including Bernal bilayer graphene (BBG) [19–23] and RMG [24–35]. These observations suggest that superconductivity in graphene emerges from a correlated electronic background, in which spin, valley, and even layer degrees of freedom play central roles. In RMG including its moiré superlattice, intrinsic flat bands and multifold spin-valley degeneracy strongly enhance interaction effects, leading to isospin-favored symmetry breaking and a hierarchy of correlated ground states, including recently discovered fractional Chern insulators and superconductivity [24–41].

Here we investigate rhombohedral hexalayer graphene (RHG) and perform ultra-low-temperature transport measurements, using optimized dual graphite gating and contact engineering. Distinct from traditional SIT, a novel insulator-superconductor transition (IST) induced by an in-plane magnetic field ($B_\parallel$) has been observed on the electron-doped side. The critical $B_\parallel$ strongly violates the Pauli limit, and symmetry-based fermiology analysis supports a spin-polarized superconductor. In addition, at zero magnetic field we identify four superconducting states in the hole-doped regime, together with orbital multiferroicity near charge neutrality point. These results establish RHG as a pristine flat-band platform in which insulating order, superconductivity, and magnetism are intimately connected, and shed new light on the microscopic mechanisms of superconductivity in correlated graphene systems.

## Hole-doped superconductivity and orbit multiferroicity

The device structure of RHG is illustrated in Fig. 1a, where RHG is equipped with both graphite top and bottom gates and encapsulated by insulating hexagonal boron nitride (hBN) as well as an extra graphite contact layer (Methods and Fig. S1). Fig. 1b shows the phase diagram of the longitudinal resistance $R_{xx}$ versus carrier density $n$ and electric displacement field $D/\varepsilon_0$ (Methods) at zero magnetic field and the base temperature $T = 10$ mK. Notably, four low-resistance states symmetric with respect to $D$ field are observed in the hole-doped regime, featuring resistance dips. In the region labeled SC1 ~ SC4, we performed differential resistance (d$V$/d$I$) measurements versus DC current ($I$) and perpendicular magnetic field ($B_\perp$), revealing a nonlinear transition peak that is regarded as a hallmark of superconductivity (Fig. 1c and Fig. S2). SC1 exhibits a critical DC current of $I_C \approx 20$ nA and a critical perpendicular magnetic field of $B_{\perp C} \approx 10$ mT (Fig. 1c). SC2 ~ SC4 show superconducting behavior akin to SC1, yet their $I_C$ and $B_{\perp C}$ are smaller (Fig. S2). The evolution of SC1 and SC2 versus $D/\varepsilon_0$ (Fig. 1d) indicates that both states are tunable by $D$ field and symmetric with respect to it, arising from the spatial inversion symmetry of RHG. The

$R_{xx}$ for SC1 shows a rapid SMT at $T_C \approx 160$ mK (Fig. 1e). Note that SC1 and SC2 exhibit a residual resistance of 40 ~ 60 Ω. The finite residual resistance of superconductivity in RMG has been reported in several recent works [33,34]. Potential explanations include insufficient electron cooling and structural inhomogeneity such as domain walls.

Similar to rhombohedral tetralayer and pentalayer graphene [42,43], RHG hosts a layer antiferromagnetic state at $n = 0$ and $D/\varepsilon_0 = 0$. While at large $D$, the system enters a layer-polarized insulating state. Between these two regimes, we also observe two symmetric bubble phases in Fig. 1b at $|D/\varepsilon_0| \approx 0.2$ V nm$^{-1}$, as well as phase boundaries that extend into regions of larger $n$ and $D$. Fig. 1f shows $B_\perp$ scans taken at different positions within the bubble phase, corresponding to the colored points in Fig. 1b. Within the bubble phase the system exhibits an anomalous Hall effect (AHE), manifested by a hysteresis in Hall resistance $R_{xy}$. As $n$ is tuned away from the bubble phase, the AHE diminishes and eventually disappears. The AHE suggests that the bubble phase is associated with orbital magnetism originating from valley polarization [31,44–46]. Moreover, sweeping $D$ field back and forth within the bubble phase exhibits hysteresis (Fig. 1g). The hysteresis observed in both $B_\perp$ field and $D$ field indicates that the bubble phase exhibits orbital multiferroicity, which is similar to the bubble phase reported in other RMG works [31,46–48]. Interestingly, when fix $n$ and sweep $D$ field back and forth across the locations of SC3 and SC4, ferroelectric hysteresis is also observed (Fig. 1h). The coexistence of superconductivity and magnetoelectric effects highlights the complex electronic interactions in the low-energy regime of RHG.

## $B_\parallel$-driven insulator-superconductor transition

We now investigate how the phase diagram of RHG responds to magnetic fields. Fig. 2a maps out the phase diagram of $R_{xx}$ at a fixed $B_\parallel$ of -0.8 T. Remarkably, in the electron-doped region around $D/\varepsilon_0 \approx -0.3$ V nm$^{-1}$, a distinct dip of $R_{xx}$ appears, marked by the dashed box and SC5 in Fig. 2a. The zoom-in phase diagram of $R_{xx}$ in Fig. 2b further shows that the dip corresponds to the previous phase boundary and extends to higher $D/\varepsilon_0$. Nonlinear differential resistance measurements performed at SC5 also reveal superconducting features (Fig. 2c). Similar to SC1 ~ SC4, the $B_\parallel$-induced SC5 also exhibits a finite residual resistance, reaching a minimum of ~ 40 Ω. Fig. 2d shows the temperature dependence of SC5 with a critical $T_C \approx 80$ mK. Fig. 2e presents the phase diagram of $R_{xx}$ versus $n$ and $D/\varepsilon_0$, where the phase boundary shifts toward lower $n$ with increasing $B_\parallel$, and the system enters the superconducting state when $B_\parallel$ reaches approximately -0.21 T. Moreover, the superconducting region is symmetric with respect to the direction of $B_\parallel$ and persists up to about 2 T (Fig. S3). We propose that the upper critical $B_\parallel$ of SC5 violates the Pauli limit, which can be predicted by weak-coupling BCS theory as $B_P \approx 1.25 \, k_B T_C/\mu_B$, where $k_B$ is the Boltzmann constant and $\mu_B$ is the Bohr magneton. Strictly speaking, the absence of SC5 at zero field ask setting $T_C$ (0 K) = 0 in the formula, leading to a predicted Pauli limit field $B_P = 0$, which is clearly ill-defined. For a qualitative comparison, we merely provide a rough estimate, $B_P \approx$ 0.15 T for $T_C = 80$ mK, to illustrate that the experimentally determined upper critical field violates the Pauli limit by orders of magnitude. By a similar estimate, SC1 ~ SC4 also show Pauli limit violation ($B_P \approx 0.3$ T for $T_C = 160$ mK), as they remain robust up to $B_\parallel \approx -0.8$ T (Fig. 2a).

Though SC5 is highly robust under $B_\parallel$, it can be rapidly suppressed by $B_\perp$. As shown in Fig. 2f, SC5 has a critical $B_\perp$ of ~ 5 mT and a critical DC current of ~ 5 nA, both of which are smaller than those of the hole-doped superconducting states. Notably, on the positive $D$ field we defined, the resistance along the phase boundary is also reduced under $B_\parallel$ but does not fully develop into a superconducting state (Fig. S4). $B_\parallel$-induced superconductivity has previously been reported in BBG [19], where it was interpreted as a spin-polarized superconductor consistent with spin-triplet pairing. SC5 observed here in RHG may share a similar origin. However, the required tuning parameters differ markedly between the two systems. In BBG, such superconducting phase appears at hole-doped regime under a large $D$ field of approximately $D/\varepsilon_0 \approx 1$ V nm$^{-1}$. By contrast, the SC5 state in RHG emerges in the electron-doped regime and requires a much smaller $D$ field of approximately -0.3 V nm$^{-1}$. The substantial reduction of the electric field makes such superconductivity more accessible in RHG experimentally.

In graphene-based systems, breaking of the spin and valley isospin symmetries can lead to changes in the Fermi surface topology. The Fermi surface geometry can be probed via high-resolution quantum oscillation measurements, an analysis method commonly employed in previous studies of BBG and RMG [19–21,24,49]. To further confirm the spin-polarized nature of SC5 and the Fermi surface topology configuration in its vicinity, we analyze the quantum oscillation data at $D/\varepsilon_0$ = -0.36 V nm$^{-1}$ and $B_\parallel$ = -0.8 T (Fig. 3a-c), where the superconducting state occurs at $n \approx 0.8 \times 10^{12}$ cm$^{-2}$. Fig. 3b show the fast Fourier transform (FFT) of quantum oscillation obtained from Fig. 3a. The vertical axis is defined as the normalized oscillation frequency $f_v = f \times e/nh$ ($f$, the oscillation frequency of $R_{xx}$ ($B_\perp$); $e$, the electron charge; $h$, Planck's constant). Clearly, the FFT results exhibit a series of peaks at certain specific values of $f_v$. Line cuts taken from Fig. 3b show the FFT of $R_{xx}$ versus $f_v$ at different fixed $n$, where the metallic states on the left and right sides of SC5 exhibit markedly different isospin polarization (Fig. 3c). At $n$ = 0.56 × 10$^{12}$ cm$^{-2}$ (the bottom panel of Fig. 3c), the FFT of $R_{xx}$ shows a peak at $f_v$ = 1/4, indicating a fourfold-degenerate Fermi surface and thus a full metal. As $n$ increases and after passing SC5 ($n$ = 0.9 × 10$^{12}$ cm$^{-2}$, the top panel of Fig. 3c), the FFT peak appears at $f_v$ = 1/2, indicating a half-metal. Combined with the previously shown $B_\parallel$ response, we identify SC5 as a spin-polarized superconductivity. In the absence of $B_\parallel$, the phase boundaries correspond to spin-unpolarized or weakly spin-polarized states and manifest as resistance peaks in $R_{xx}$ (Fig. 3d). Fig. 3e and f show the temperature dependence of the resistance peak at $B_\parallel$ = 0. Surprisingly, although the resistance peak at the phase boundary is only about 600 Ω, it exhibits clear insulating behavior, with a thermally activated gap of $\Delta \approx 0.1$ meV. Under increasing $B_\parallel$, the insulating gap at the phase boundary is suppressed, and the system gradually enters the superconducting state. While with increasing $B_\perp$, the resistance peak gradually decreases and eventually evolves into the quantum Hall regime with the formation of Landau levels (Fig. S3d).

## Discussion

Such IST on the electron-doped side driven by $B_\parallel$ has not been previously reported in other systems, nor has it been theoretically predicted. In conventional superconductors, an external magnetic field typically suppresses superconductivity, driving the system into a metallic or insulating state. In stark contrast, our experiments reveal that $B_\parallel$ induces a transition from insulator to superconductor. This anomalous behavior suggests that this insulating phase is not a trivial insulator but involves spin- or isospin-related correlations that can be selectively tuned by $B_\parallel$. Meanwhile, the distinct roles of $B_\parallel$ and $B_\perp$ fields are remarkable. $B_\parallel$ primarily couples to electron spins via the Zeeman effect, with minimal orbital disruption. This spin polarization may relieve constraints in the insulating state and activate spin-polarized triplet pairing channels, enabling $B_\parallel$-induced superconductivity. The observed SC5 with significant Pauli-limit violation, together with the location on the spin-polarized side of the phase boundary in the fermiology analysis, support its identification as a spin-polarized superconductor primarily governed by a triplet-pairing mechanism. By contrast, $B_\perp$ field introduces strong orbital effects that break Cooper pairs, suppressing superconductivity. Unlike SC5, although SC1 ~ SC4 also violate the Pauli limit, their response to the $B_\parallel$ just remains robust without inducement or enhancement effect. Plausible explanations for paring mechanism of SC1 ~ SC4 may include an admixture of spin-singlet and -triplet components [50–53] or some compensation mechanism [54–56]. The details for pairing mechanism of these superconducting states in RHG remain unclear and call for further theoretical and experimental investigation.

In conclusion, we have demonstrated a wealth of superconductivity on both the electron- and hole-doped sides in RHG. Particularly, we have observed a novel IST driven by $B_\parallel$. Our findings establish isospin symmetry breaking as a central organizing principle governing both insulating and pairing mechanism in RHG, and identify the $B_\parallel$ as a powerful tuning parameter that selectively stabilizes superconductivity while favoring spin polarization. These results expand the landscape of field-enabled superconductivity and underscore RMG as a clean and versatile platform for exploring potential unconventional superconductivity beyond the BCS framework.

*Note added*: We noted several relevant studies showing superconductivity induced by $B_\parallel$ on the hole-doped side during the preparation of our manuscript [27,33,34].

## Acknowledgements

X.B.L. acknowledges support from the National Key R&D Program (Grant Nos. 2022YFA1403500 and 2024YFA1409002) and the National Natural Science Foundation of China (Grant Nos. 12274006 and 12141401). K.W. and T.T. acknowledge support from the JSPS KAKENHI (Grant Nos. 21H05233 and 23H02052) and World Premier International Research Center Initiative (WPI), MEXT, Japan. J.X. acknowledges support from the China Association for Science and Technology Youth Talent Promotion Program.

## Author Contributions

X.B.L., Z.H. and J.X. conceived and designed the experiments; J.X. and Z.H. fabricated the devices and performed the transport measurement with help from Z.C., Z.Z. and X.L.; K.W. and T.T. provided the hBN crystals; J.X., Z.H., and X.B.L. wrote the paper with input from others.

## Competing interests

The authors declare no competing interests.

## Data Availability

All data supporting the findings of this study are available within the main text, figures and Supplementary Materials, or from the corresponding authors upon request.

## Appendix: Methods

**Device fabrication.** The RHG device fabrication and basic characterization mostly follows protocols illustrated in our previous work [38]. The preliminary preparation involved: mechanical exfoliation of graphene onto the 285 nm $SiO_2$/Si substrate, thickness calibration by optical contrast, stacking configuration determination via rapid infrared imaging technique and Raman spectroscopy, and shape definition by laser cutting. Specially, the high-resistance state of RHG at the charge neutrality point can lead to poor electrical contacts. To mitigate this impact on transport measurements, we employed extra graphene flakes as the side graphene contacts, eliminating the need for large silicon back-gate voltages. Using a fry-transfer technique, we sequentially picked up the top graphene gate, top hBN, side graphene contacts and RHG samples, and then placed the stack onto the bottom hBN and gate substrate that had been pre-cleaned by annealing in $Ar/H_2$ (9:1) mixture atmosphere at 350 °C followed by mechanical cleaning via AFM contact-mode scanning. The device was sculpted into Hall bar geometries through standard electron beam lithography and reactive ion etching, with edge contacts formed by evaporated Cr/Au (5/60nm) electrodes.

**Transport measurement.** The transport measurement was operated by standard low-frequency lock-in techniques. The low-temperature environment was provided by an Oxford Instruments dilution refrigerator unless otherwise specified, which performed at its base phonon temperature of approximately 10mK. The four-terminal transport measurements were performed utilizing SR860 lock-in amplifiers, cascaded with an SR560 voltage preamplifier to enhance the signal-to-noise ratio and using an AC current bias ranging from 0.5 to 2nA at a frequency of 17.777Hz. Dual-gate voltage was applied by two Keithley 2400 source-meters.

The dual gate structure allows independent control of the carrier density $n = (C_{tg}V_{tg} + C_{bg}V_{bg})/e$ and the electric displacement field $D/\varepsilon_0 = (C_{tg}V_{tg} - C_{bg}V_{bg})/(2\varepsilon_0)$, where $\varepsilon_0$ is the permittivity of vacuum, and $C_{tg}$ and $C_{bg}$ are the capacitances of the top and bottom gates per unit area, respectively.

During the high $B_\parallel$ measurements, an additional small $B_\perp$ was applied to compensate the residual out-of-plane component at high $B_\parallel$. Therefore, the suppression of superconductivity is caused by the $B_\parallel$, rather than by any perpendicular component.

# References

[1] J. Bardeen, L. N. Cooper, and J. R. Schrieffer, Theory of Superconductivity, Phys. Rev. **108**, 1175 (1957).
[2] J. Bardeen, L. N. Cooper, and J. R. Schrieffer, Microscopic Theory of Superconductivity, Phys. Rev. **106**, 162 (1957).
[3] A. M. Clogston, Upper Limit for the Critical Field in Hard Superconductors, Phys. Rev. Lett. **9**, 266 (1962).
[4] M. P. A. Fisher, Quantum phase transitions in disordered two-dimensional superconductors, Phys. Rev. Lett. **65**, 923 (1990).
[5] A. M. Finkel'stein, Suppression of superconductivity in homogeneously disordered systems, Phys. B Condens. Matter **197**, 636 (1994).
[6] N. Marković, C. Christiansen, A. M. Mack, W. H. Huber, and A. M. Goldman, Superconductor-insulator transition in two dimensions, Phys. Rev. B **60**, 4320 (1999).
[7] B. Sacépé, M. Feigel'man, and T. M. Klapwijk, Quantum breakdown of superconductivity in low-dimensional materials, Nat. Phys. **16**, 734 (2020).

[8] S. S. Saxena et al., Superconductivity on the border of itinerant-electron ferromagnetism in UGe2, Nature **406**, 587 (2000).

[9] D. Aoki, A. Huxley, E. Ressouche, D. Braithwaite, J. Flouquet, J.-P. Brison, E. Lhotel, and C. Paulsen, Coexistence of superconductivity and ferromagnetism in URhGe, Nature **413**, 613 (2001).

[10] D. Aoki and J. Flouquet, Ferromagnetism and Superconductivity in Uranium Compounds, J. Phys. Soc. Jpn. **81**, 011003 (2012).

[11] A. M. Mounce, H. Yasuoka, G. Koutroulakis, N. Ni, E. D. Bauer, F. Ronning, and J. D. Thompson, Detection of a Spin-Triplet Superconducting Phase in Oriented Polycrystalline $U_2PtC_2$ Samples Using $Pt^{195}$ Nuclear Magnetic Resonance, Phys. Rev. Lett. **114**, 127001 (2015).

[12] S. Ran et al., Nearly ferromagnetic spin-triplet superconductivity, Science **365**, 684 (2019).

[13] S. Ran et al., Extreme magnetic field-boosted superconductivity, Nat. Phys. **15**, 1250 (2019).

[14] L. Jiao, S. Howard, S. Ran, Z. Wang, J. O. Rodriguez, M. Sigrist, Z. Wang, N. P. Butch, and V. Madhavan, Chiral superconductivity in heavy-fermion metal UTe2, Nature **579**, 523 (2020).

[15] D. Aoki, J.-P. Brison, J. Flouquet, K. Ishida, G. Knebel, Y. Tokunaga, and Y. Yanase, Unconventional superconductivity in $UTe_2$, J. Phys. Condens. Matter **34**, 243002 (2022).

[16] K. Ishida, H. Mukuda, Y. Kitaoka, K. Asayama, Z. Q. Mao, Y. Mori, and Y. Maeno, Spin-triplet superconductivity in Sr2RuO4 identified by 17O Knight shift, Nature **396**, 658 (1998).

[17] G. M. Luke et al., Time-reversal symmetry-breaking superconductivity in Sr2RuO4, Nature **394**, 558 (1998).

[18] K. D. Nelson, Z. Q. Mao, Y. Maeno, and Y. Liu, Odd-Parity Superconductivity in $Sr_2RuO_4$, Science **306**, 1151 (2004).

[19] H. Zhou, L. Holleis, Y. Saito, L. Cohen, W. Huynh, C. L. Patterson, F. Yang, T. Taniguchi, K. Watanabe, and A. F. Young, Isospin magnetism and spin-polarized superconductivity in Bernal bilayer graphene, Science **375**, 774 (2022).

[20] Y. Zhang, R. Polski, A. Thomson, É. Lantagne-Hurtubise, C. Lewandowski, H. Zhou, K. Watanabe, T. Taniguchi, J. Alicea, and S. Nadj-Perge, Enhanced superconductivity in spin–orbit proximitized bilayer graphene, Nature **613**, 268 (2023).

[21] C. Li et al., Tunable superconductivity in electron- and hole-doped Bernal bilayer graphene, Nature **631**, 300 (2024).

[22] Y. Zhang et al., Twist-programmable superconductivity in spin–orbit-coupled bilayer graphene, Nature **641**, 625 (2025).

[23] L. Holleis et al., Nematicity and orbital depairing in superconducting Bernal bilayer graphene, Nat. Phys. **21**, 444 (2025).

[24] H. Zhou, T. Xie, T. Taniguchi, K. Watanabe, and A. F. Young, Superconductivity in rhombohedral trilayer graphene, Nature **598**, 434 (2021).

[25] C. L. Patterson et al., Superconductivity and spin canting in spin–orbit-coupled trilayer graphene, Nature **641**, 632 (2025).

[26] J. Yang et al., Impact of spin–orbit coupling on superconductivity in rhombohedral graphene, Nat. Mater. **24**, 1058 (2025).

[27] J. Yang et al., Magnetic Field-Enhanced Graphene Superconductivity with Record Pauli-Limit Violation, arXiv:2510.10873.

[28] T. Han et al., Signatures of chiral superconductivity in rhombohedral graphene, Nature **643**, 654 (2025).

[29] Y. Choi et al., Superconductivity and quantized anomalous Hall effect in rhombohedral graphene, Nature **639**, 342 (2025).
[30] J. Seo et al., Family of Unconventional Superconductivities in Crystalline Graphene, arXiv:2509.03295.
[31] J. Deng, J. Xie, H. Li, T. Taniguchi, K. Watanabe, J. Shan, K. F. Mak, and X. Liu, Superconductivity and Ferroelectric Orbital Magnetism in Semimetallic Rhombohedral Hexalayer Graphene, arXiv:2508.15909.
[32] R. Q. Nguyen, H.-T. Wu, E. Morissette, N. J. Zhang, P. Qin, K. Watanabe, T. Taniguchi, A. W. Hui, D. E. Feldman, and J. I. A. Li, A Hierarchy of Topological and Superconducting States in Rhombohedral Hexalayer Graphene, arXiv:2507.22026.
[33] M. Kumar, D. Waleffe, A. Okounkova, R. Tejani, K. Watanabe, T. Taniguchi, É. Lantagne-Hurtubise, J. Folk, and M. Yankowitz, Pervasive Spin-Triplet Superconductivity in Rhombohedral Graphene, arXiv:2511.16578.
[34] Y. Guo et al., Flat Band Surface State Superconductivity in Thick Rhombohedral Graphene, arXiv:2511.17423.
[35] E. Morissette, P. Qin, H.-T. Wu, N. J. Zhang, R. Q. Nguyen, K. Watanabe, T. Taniguchi, and J. I. A. Li, Striped Superconductor in Rhombohedral Hexalayer Graphene, arXiv:2504.05129.
[36] Z. Lu, T. Han, Y. Yao, A. P. Reddy, J. Yang, J. Seo, K. Watanabe, T. Taniguchi, L. Fu, and L. Ju, Fractional quantum anomalous Hall effect in multilayer graphene, Nature **626**, 759 (2024).
[37] Z. Lu et al., Extended quantum anomalous Hall states in graphene/hBN moiré superlattices, Nature **637**, 1090 (2025).
[38] J. Xie et al., Tunable fractional Chern insulators in rhombohedral graphene superlattices, Nat. Mater. **24**, 1042 (2025).
[39] S. H. Aronson, T. Han, Z. Lu, Y. Yao, J. P. Butler, K. Watanabe, T. Taniguchi, L. Ju, and R. C. Ashoori, Displacement Field-Controlled Fractional Chern Insulators and Charge Density Waves in a Graphene/hBN Moiré Superlattice, Phys. Rev. X **15**, 031026 (2025).
[40] J. Xie et al., Unconventional Orbital Magnetism in Graphene-Based Fractional Chern Insulators, arXiv:2506.01485.
[41] Z. Huo et al., Does Moire Matter? Critical Moire Dependence with Quantum Fluctuations in Graphene Based Integer and Fractional Chern Insulators, arXiv:2510.15309.
[42] K. Liu, Spontaneous broken-symmetry insulator and metals in tetralayer rhombohedral graphene, Nat. Nanotechnol. **19**, (2024).
[43] T. Han, Z. Lu, G. Scuri, J. Sung, J. Wang, T. Han, K. Watanabe, T. Taniguchi, H. Park, and L. Ju, Correlated insulator and Chern insulators in pentalayer rhombohedral-stacked graphene, Nat. Nanotechnol. **19**, 181 (2024).
[44] Y. Sha, J. Zheng, K. Liu, H. Du, K. Watanabe, T. Taniguchi, J. Jia, Z. Shi, R. Zhong, and G. Chen, Observation of a Chern insulator in crystalline ABCA-tetralayer graphene with spin-orbit coupling, Science **384**, 414 (2024).
[45] T. Han et al., Large quantum anomalous Hall effect in spin-orbit proximitized rhombohedral graphene, Science **384**, 647 (2024).
[46] K. Krötzsch, A. Herasymchuk, Y. Zhumagulov, A. Magrez, K. Watanabe, T. Taniguchi, S. G. Sharapov, O. V. Yazyev, and M. Banerjee, Magnetoelectric Switching of Magnetic Order in Rhombohedral Graphene, arXiv:2509.24672.
[47] T. Han et al., Orbital multiferroicity in pentalayer rhombohedral graphene, Nature **623**, 41 (2023).

[48] E. Morissette, P. Qin, K. Watanabe, T. Taniguchi, and J. I. A. Li, Evidence of Momentum Space Condensation in Rhombohedral Hexalayer Graphene, arXiv:2503.09954.

[49] H. Zhou et al., Half- and quarter-metals in rhombohedral trilayer graphene, Nature **598**, 429 (2021).

[50] J. M. Lu, O. Zheliuk, I. Leermakers, N. F. Q. Yuan, U. Zeitler, K. T. Law, and J. T. Ye, Evidence for two-dimensional Ising superconductivity in gated $MoS_2$, Science **350**, 1353 (2015).

[51] X. Xi, Z. Wang, W. Zhao, J.-H. Park, K. T. Law, H. Berger, L. Forró, J. Shan, and K. F. Mak, Ising pairing in superconducting $NbSe_2$ atomic layers, Nat. Phys. **12**, 139 (2016).

[52] B. T. Zhou, N. F. Q. Yuan, H.-L. Jiang, and K. T. Law, Ising superconductivity and Majorana fermions in transition-metal dichalcogenides, Phys. Rev. B **93**, 180501 (2016).

[53] D. Möckli and M. Khodas, Magnetic-field induced s + if pairing in Ising superconductors, Phys. Rev. B **99**, 180505 (2019).

[54] V. Jaccarino and M. Peter, Ultra-High-Field Superconductivity, Phys. Rev. Lett. **9**, 290 (1962).

[55] P. Fulde and R. A. Ferrell, Superconductivity in a Strong Spin-Exchange Field, Phys. Rev. **135**, A550 (1964).

[56] H. W. Meul, C. Rossel, M. Decroux, Ø. Fischer, G. Remenyi, and A. Briggs, Observation of Magnetic-Field-Induced Superconductivity, Phys. Rev. Lett. **53**, 497 (1984).

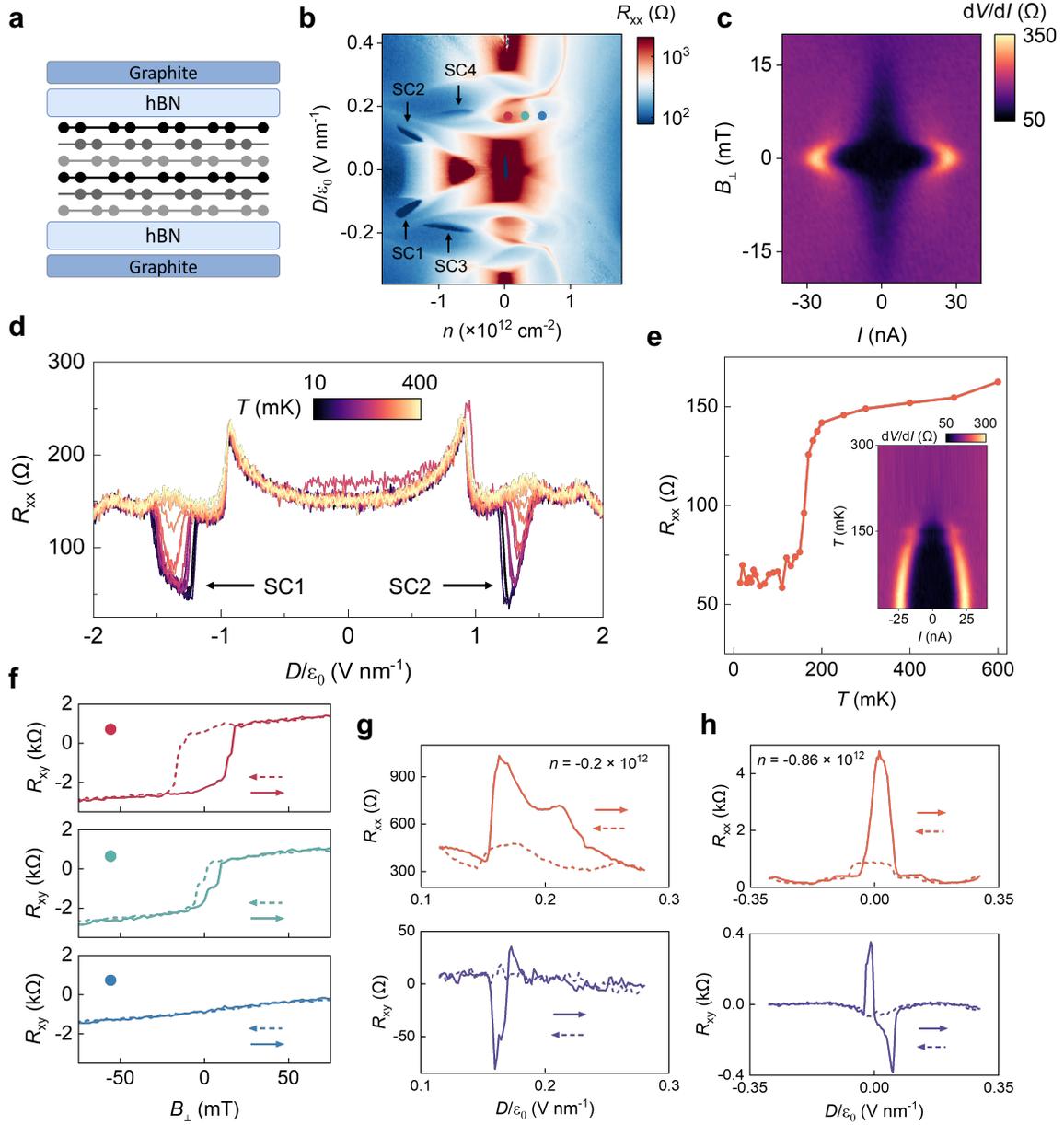

**FIG. 1. Hole-doped superconductivity and orbit multiferroicity. a,** Schematic of the device configuration. **b,** Phase diagram of $R_{xx}$ as functions of $n$ and $D/\varepsilon_0$ measured at zero magnetic field. **c,** Differential resistance $dV/dI$ versus DC current $I$ and $B_\perp$ field taken from SC1 at $n = -1.49 \times 10^{12}$ cm$^{-2}$ and $D/\varepsilon_0 = -0.13$ V nm$^{-1}$. **d,** Line cuts of $R_{xx}$ versus $D/\varepsilon_0$ measured at different temperatures and fixed $n = -1.49 \times 10^{12}$ cm$^{-2}$. **e,** Temperature dependence of $R_{xx}$ measured at SC1, featuring $T_C \approx 160$ mK. The insert shows the $dV/dI$ map versus $I$ and $T$. **f,** $B_\perp$-field scans of $R_{xy}$ obtained at $D/\varepsilon_0 = 0.17$ V nm$^{-1}$, corresponding to the bubble phase in **b**. Colored points denote different locations in the bubble phase. **g,** $D$-field scans of $R_{xx}$ (top panel) and $R_{xy}$ (bottom panel) at $n = -0.2 \times 10^{12}$ cm$^{-2}$ in the bubble phase, showing ferroelectric hysteresis. **h,** Sweeping $D$-field back and forth at $n = -0.86 \times 10^{12}$ cm$^{-2}$, crossing SC3 and SC4. Both $R_{xx}$ (top panel) and $R_{xy}$ (bottom panel) exhibit ferroelectric hysteresis emerged near $D/\varepsilon_0 = 0$.

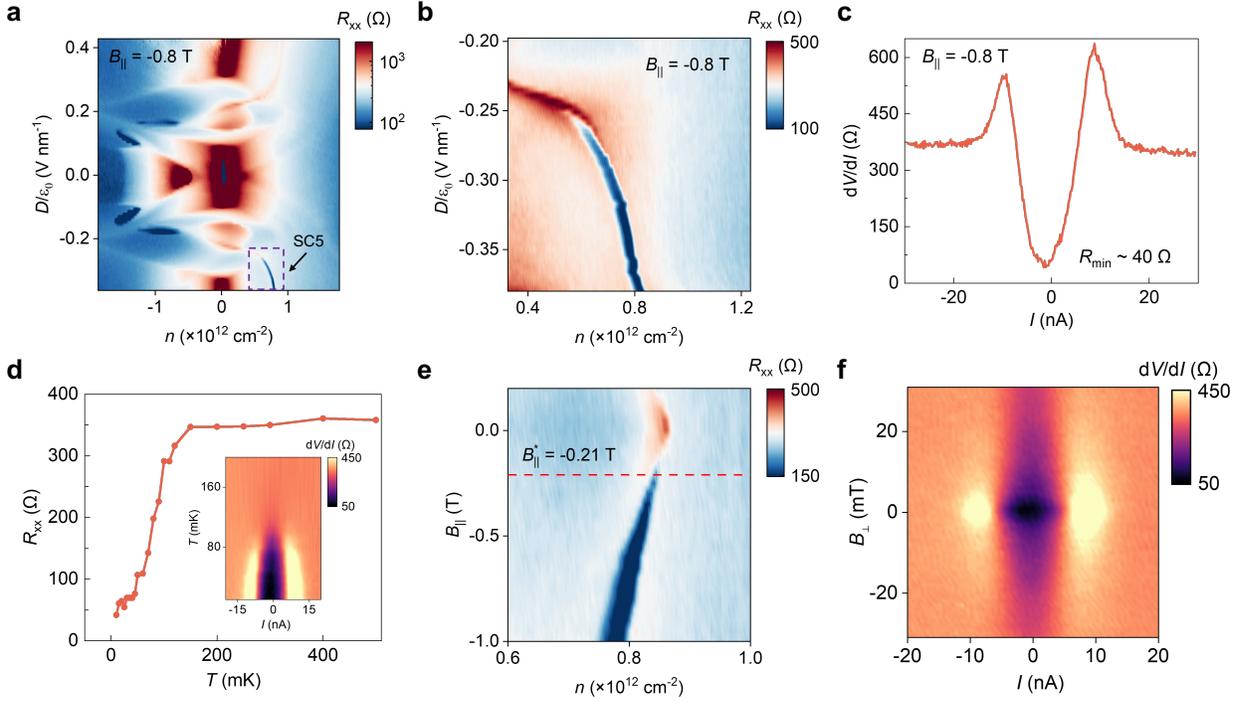

**FIG. 2.** $B_\parallel$-**induced superconductivity. a,** Phase diagram of $R_{xx}$ versus $n$ and $D/\varepsilon_0$ obtained at $B_\parallel$ = -0.8 T. **b,** Zoomed-in map of $R_{xx}$ in the $n$-$D/\varepsilon_0$ parameter space, corresponding to the dashed box labeled SC5 in **a**. At $D/\varepsilon_0 \approx$ -0.26 V nm$^{-1}$, the phase boundary develops a dip and evolves along the boundary toward higher $D$ fields. **c,** $dV/dI$ versus $I$ obtained at $n = 0.78 \times 10^{12}$ cm$^{-2}$ and $D/\varepsilon_0$ = -0.36 V nm$^{-1}$, featuring a nonlinear transition peak of SC5. The minimum resistance is about 40 Ω. **d,** Temperature dependence of SC5. $R_{xx}$ shows a rapid drop around 80 mK. The insert shows the $dV/dI$ map versus $I$ and $T$. **e,** Phase diagram of $R_{xx}$ versus $n$ and $B_\parallel$ measured at $D/\varepsilon_0$ = -0.36 V nm$^{-1}$. At a critical field of $B_\parallel^* \approx$ -0.21 T, the phase boundary evolves from a resistive state into a superconductor and persists up to high $B_\parallel \approx$ 2 T (Fig. S3a and Fig. S3b). **f,** $dV/dI$ as functions of $I$ and $B_\perp$ at fixed $B_\parallel$ = -0.8 T, showing a critical $B_\perp$ of 5 mT.

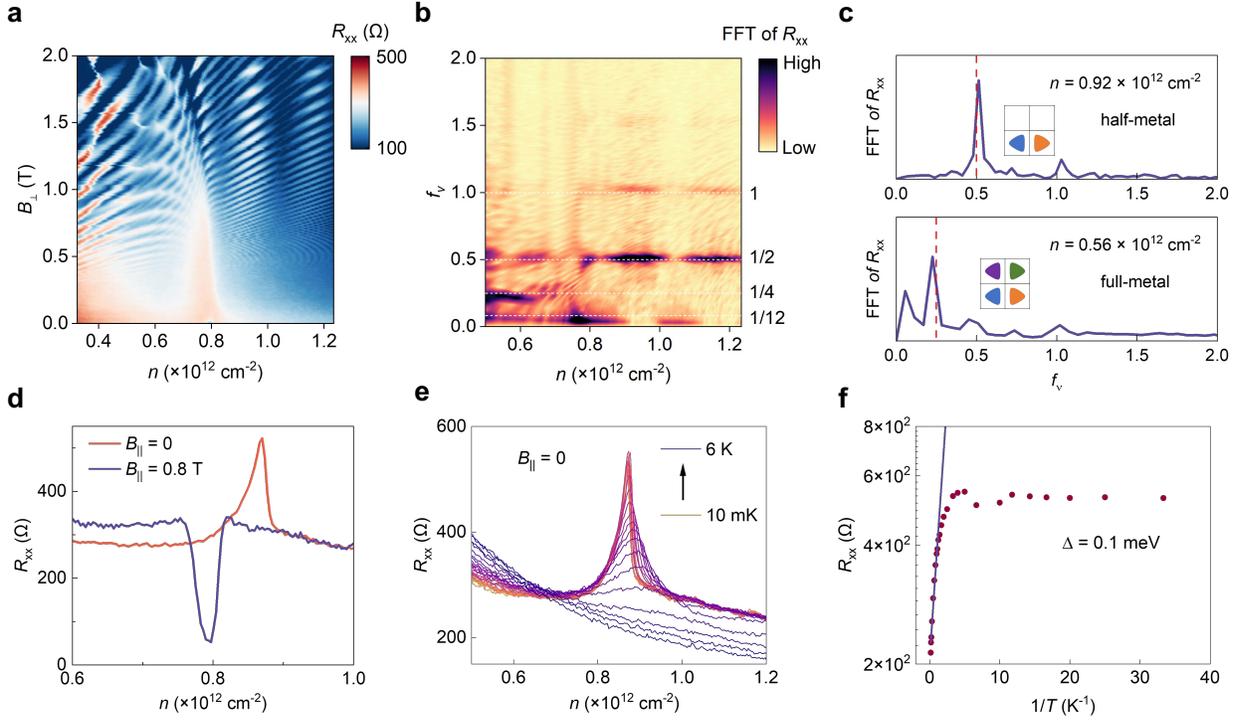

**FIG. 3. Insulator-superconductor transition. a,** Phase diagram of $R_{xx}$ versus $n$ and $B_\perp$ obtained at $B_\parallel = -0.8$ T and $D/\varepsilon_0 = -0.36$ V nm$^{-1}$, exhibiting distinct quantum oscillations on the two sides of the phase boundary. **b,** Fourier transform of $R_{xx}(1/B_\perp)$, calculated from the data in **a**. The vertical axis is defined as the normalized oscillation frequency $f_v = f \times e/nh$. $f$, the oscillation frequency of $R_{xx}(B_\perp)$; $e$, the electron charge; $h$, Planck's constant. Note that only data within 0.7 T $< B_\perp <$ 1.5 T are used in the calculation. **c,** Line cuts of FFT taken from the map in **b** at $n = 0.92 \times 10^{12}$ cm$^{-2}$ (top panel) and $n = 0.56 \times 10^{12}$ cm$^{-2}$ (bottom panel). The inset schematic illustrates the possible isospin polarization and Fermi surface configuration. **d,** $R_{xx}$ versus $n$ obtained at $B_\parallel = 0$ and -0.8 T, with fixed $D/\varepsilon_0 = -0.36$ V nm$^{-1}$. The phase boundary evolves into superconductivity and favors low carrier doping under finite $B_\parallel$. **e,** Temperature dependence of the phase boundary at $B_\parallel = 0$ and $D/\varepsilon_0 = -0.36$ V nm$^{-1}$. **f,** Plot of $R_{xx}$ versus $1/T$ and estimate of the thermally activated gap ($\Delta$) of the phase boundary at $B_\parallel = 0$ and $D/\varepsilon_0 = -0.36$ V nm$^{-1}$. Solid line is the linear fit to $R_0 e^{-\Delta/k_B T}$, where $R_0$ is the fitting constant and $k_B$ is the Boltzmann constant.

Supplemental Materials for

# Magnetic-Field-Driven Insulator-Superconductor Transition in Rhombohedral Graphene

Jian Xie[1†], Zihao Huo[1†]*, Zhimou Chen[1], Zaizhe Zhang[1], Kenji Watanabe[2], Takashi Taniguchi[3], Xi Lin[1,4,5]* and Xiaobo Lu[1,6]*

[1]*International Center for Quantum Materials, School of Physics, Peking University, Beijing 100871, China*
[2]*Research Center for Electronic and Optical Materials, National Institute for Material Sciences, 1-1 Namiki, Tsukuba 305-0044, Japan*
[3]*Research Center for Materials Nanoarchitectonics, National Institute for Material Sciences, 1-1 Namiki, Tsukuba 305-0044, Japan*
[4]*Hefei National Laboratory, Hefei 230088, China*
[5]*Interdisciplinary Institute of Light-Element Quantum Materials and Research Center for Light-Element Advanced Materials, Peking University, Beijing 100871, China*
[6]*Collaborative Innovation Center of Quantum Matter, Beijing 100871, China*

[†]These authors contributed equally to this work.
*E-mails: xiaobolu@pku.edu.cn; xilin@pku.edu.cn; zhhuo@stu.pku.edu.cn

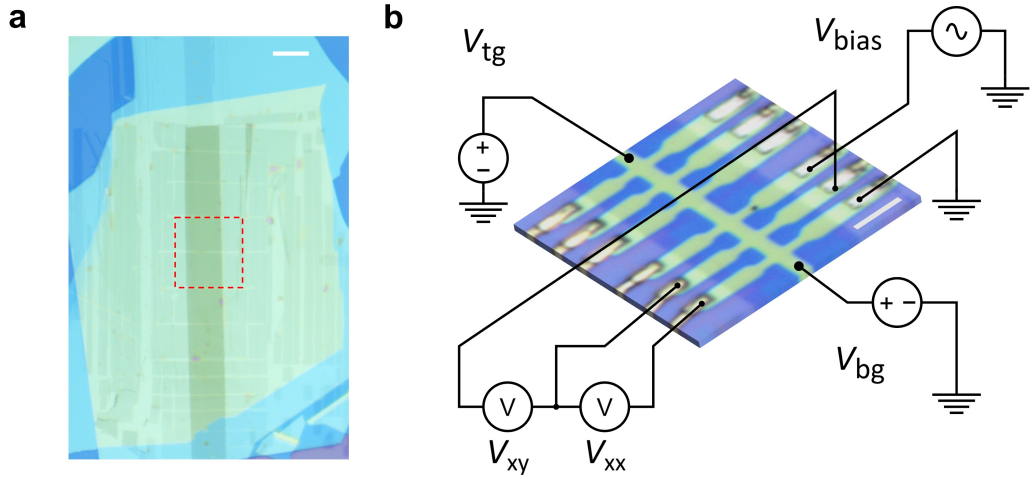

**FIG. S1. Characterization of the device and measurement configurations. a,** Optical image of the RHG sample. The red dashed box represents the sample regions we fabricated. Scale bar, 10 μm. **b,** Schematic diagram of the four-wire electrical measurement. Scale bar, 4 μm. $V_{tg}$, top gate voltage; $V_{bg}$, bottom gate voltage; $V_{xx}$, longitudinal voltage; $V_{xy}$, Hall voltage; $V_{bias}$, bias voltage.

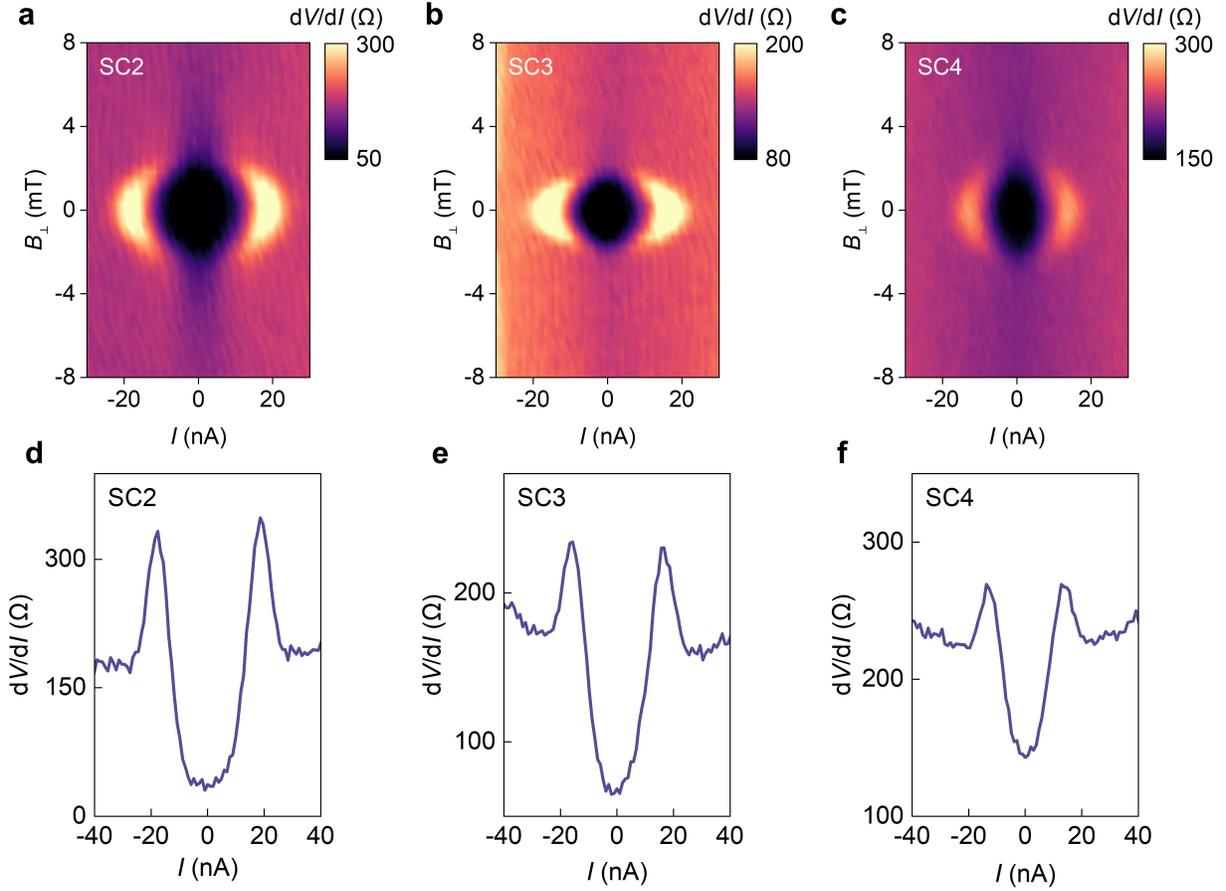

FIG. S2. **Identification of the superconductivity in SC2 ~ SC4. a-c,** Differential resistance d$V$/d$I$ versus DC current $I$ and perpendicular magnetic field $B_\perp$ obtained at SC2 ($n = -0.77 \times 10^{12}$ cm$^{-2}$, $D/\varepsilon_0 = -0.11$ V nm$^{-1}$) (**a**), SC3 ($n = -0.81 \times 10^{12}$ cm$^{-2}$, $D/\varepsilon_0 = -0.18$ V nm$^{-1}$) (**b**) and SC4 ($n = -0.61 \times 10^{12}$ cm$^{-2}$, $D/\varepsilon_0 = -0.18$ V nm$^{-1}$) (**c**), respectively. **d-f,** Line cuts of d$V$/d$I$ versus $I$ obtained at SC2 (**d**), SC3 (**e**) and SC4 (**f**), showing sharp superconducting transitions near the critical current. The data in **d-f** are taken from the maps of **a-c**, respectively.

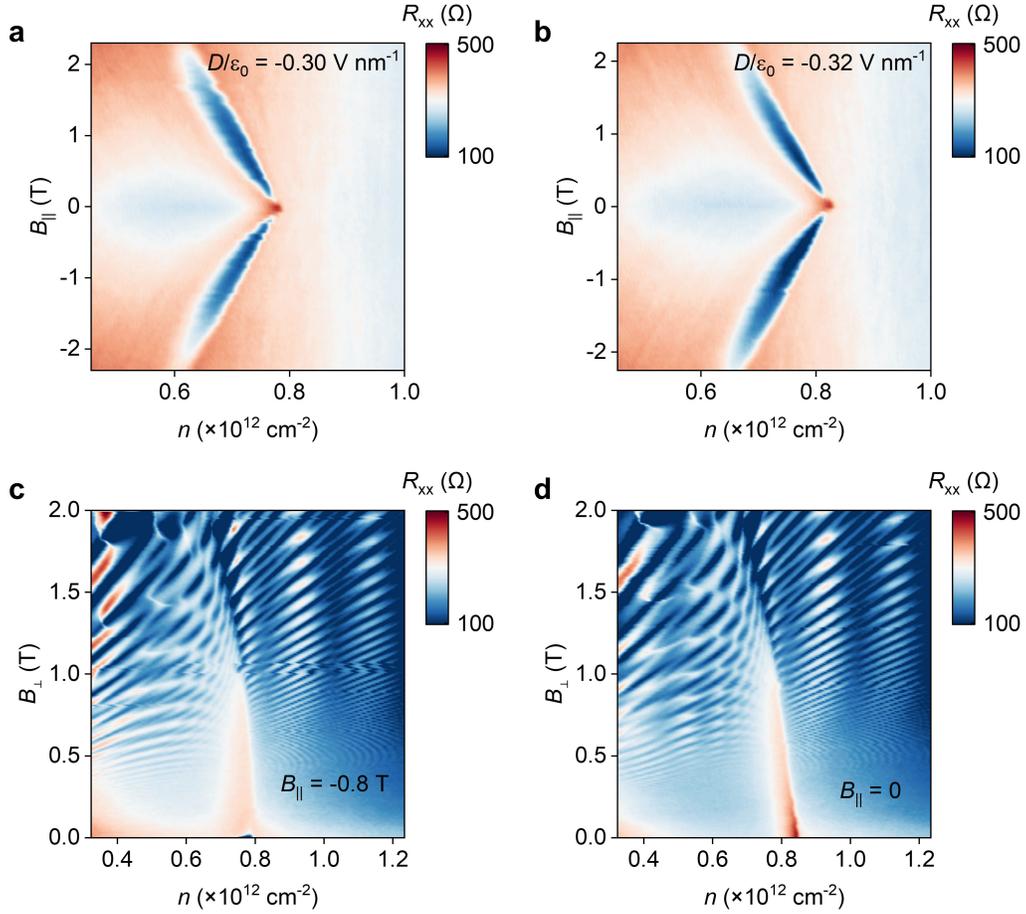

**FIG. S3. Evolution of the phase boundary in magnetic fields. a,b,** Phase diagram of $R_{xx}$ versus $n$ and $B_\parallel$ measured at $D/\varepsilon_0$ = -0.30 V nm$^{-1}$ (**a**) and -0.32 V nm$^{-1}$ (**b**), respectively. Similar to the behavior shown in Fig. 2e, superconductivity is induced by a small $B_\parallel$ and can persists up to approximately 2 T. Note that an additional $B_\perp$ is applied to compensate for the residual perpendicular component at large $B_\parallel$. **c,d,** Landau fan diagram of $R_{xx}$ as functions of $n$ and $B_\perp$ measured at $D/\varepsilon_0$ = -0.33 V nm$^{-1}$ with fixed $B_\parallel$ = -0.8 T (**c**) and 0 (**d**), respectively.

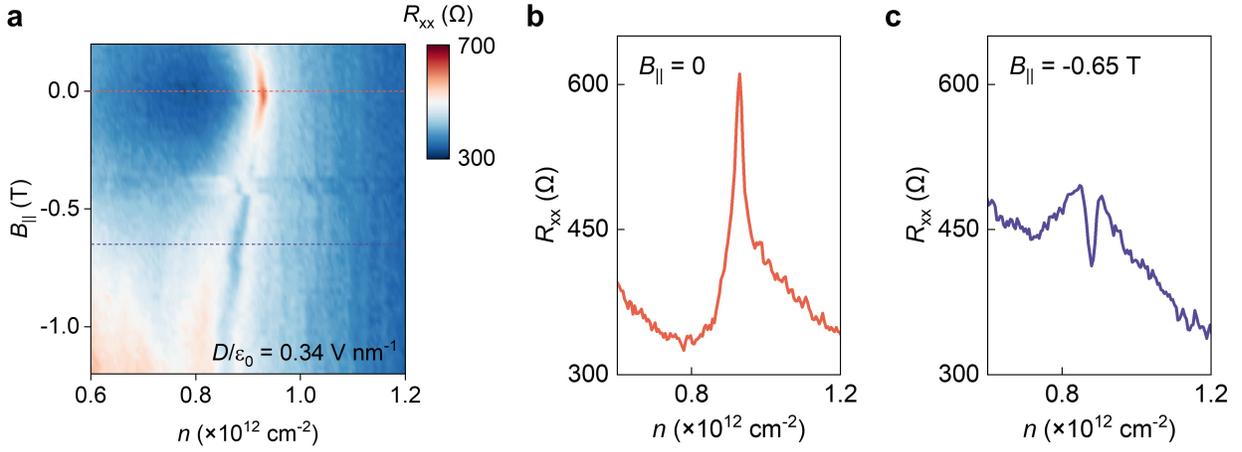

**FIG. S4. Incomplete development of superconductivity on the positive $D$ field. a,** Phase diagram of $R_{xx}$ versus $n$ and $B_\parallel$ measured at $D/\varepsilon_0 = 0.34$ V nm$^{-1}$, showing spin-polarized features similar to those observed on the negative $D$ field side. **b,c,** Line cuts of $R_{xx}$ versus $n$ corresponding to the dashed line in **a**. At $B_\parallel = 0$ (**b**), the phase boundary behaves as a spin-unpolarized insulator. Upon applying a finite $B_\parallel$ of -0.65 T (**c**), the resistance peak evolves into a dip, but does not fully develop into a superconducting state.